\documentclass[12pt]{iopart}
\usepackage{graphicx}

\begin{document}

\title{Configurational entropy of hard spheres}

\author{Luca Angelani$^1$ and \ Giuseppe Foffi$^{2}$}

\address{$^1$ Research center SMC INFM-CNR, c/o Universit\`a
di Roma ``La Sapienza'', P.le A. Moro 2, I-00185, Roma, Italy}

\address{$^2$ Institut Romand de Recherche Num\'erique en
Physique des Mat\'eriaux IRRMA,
 and Institute of Theoretical Physics (ITP), Ecole Polytechnique F\'ed\'erale de Lausanne (EPFL), CH-1015 Lausanne, Switzerland
}

\begin{abstract}
We  numerically calculate the configurational entropy $S_{conf}$ of a binary mixture 
of hard spheres, by using a perturbed Hamiltonian method trapping the system inside a given state,
which requires less assumptions than the previous methods
[R.J.~Speedy, Mol. Phys. {\bf 95}, 169 (1998)].
We find that  $S_{conf}$ is  a decreasing function of  packing fraction $\varphi$
and extrapolates to zero at the Kauzmann packing fraction
$\varphi_{_K} \simeq 0.62$, suggesting the possibility of an ideal 
glass-transition for hard spheres system.
Finally, the Adam-Gibbs relation is found to hold.
\end{abstract}

\section{Introduction}

The idea that the glass transition is driven by a decreasing of the number of
accessible states upon lowering temperature (or raising density) is quite old
\cite{kauzmann,AG,GM}.  In this picture,
if crystallization is avoided,
an ideal glass transition is expected
to happen at the point where the configurational entropy $S_{conf}$ (the
logarithm of the number of states) vanishes.  When the liquid enters in the
the supercooled region, the dynamics becomes slower and slower and the
particles get trapped for an increasingly longer time
 inside the ``cages'' made by their
neighbors: the dynamics of the system can be successfully described as a
``fast'' motion of the representative point in the $3N$ configuration space
inside metastable states, and a ``slow'' motion corresponding to jumps among
states. Entering more in the supercooled region the number of accessible
metastable states decreases
and the extrapolation to zero of $S_{conf}$ defines the ideal glass
transition.  In experiments (or numerical simulations) the region close to the
ideal glass transition is unreachable, due to the ``apparent'' arrest of the
system at the so called glass-transition temperature (or density) when
relaxation times become longer than experimental time scale.  The above
scenario has been shown to be valid for many interacting systems, based on
smooth pair-potential (as Lennard-Jones liquids), for which the Potential
Energy Landscape (PEL) approach
\cite{deb_nature,stilweb,stil,skt,scala,land_sastry,land_buc,land_sastry2,keyes_inm,fsschool}
and the replica method \cite{mezard,coluzzi0} have allowed to give numerical
estimations of $S_{conf}$ and of the ideal glass transition.

The overall picture is still not well established for Hard Spheres (HS), for
which the existence of a glass transition is still an
open question \cite{rt,robles,speedy3,tarzia,donev}.
A particularly important role seems to be played by the dimesionality of the
system. In particular in d=2 dimension there are numerical \cite{donev,sen_kra}
and theoretical \cite{tarzia2,zam_and}
evidences of the absence of a thermodynamic glass transition, while the
opposite seems to be true in d=3~\cite{speedy3,paza}.  
Moreover, the step-wise
form of the interparticle potential does not allow a PEL analysis 
and different
approaches have to be taken in consideration in order to calculate the
configurational entropy $S_{conf}$.  In the past, attempts to estimate
$S_{conf}$ have been performed based on different evaluations of the entropy
in each single state \cite{speedy3,speedy2,aste}.  Recently the replica method
has been extended to the HS case for one-component systems \cite{zam_and,paza}.

In this paper we  follow an approach, based on the Frenkel-Ladd method
\cite{frenkel_book} and recently introduced in the study of
Lennard-Jones systems \cite{coluzzi} and attractive colloids
\cite{noi,moreno}, to numerically estimate $S_{conf}$ for binary hard
spheres.  As in previous studies, the calculation is reduced to that
of vibrational entropy $S_{vib}$, using the fact that the total
entropy $S$ can be decomposed into the sum of a configurational
contribution $S_{conf}$ and a vibrational one $S_{vib}$:
\begin{equation}
S = S_{conf} + S_{vib} \ .
\label{entropy}
\end{equation}
This expression is consistent with the idea that, at high enough density,
there are two well separated time scales: a fast one, related to the motion
inside a local state (the rattling in the cage), and a slow one associated to 
the exploration of different states.\\
The total entropy $S$ is obtained by thermodynamic integration, starting from
the ideal gas state. The quantity $S_{vib}$ is calculated using a perturbed
Hamiltonian, adding to the original Hamiltonian an harmonic potential around a
given reference configuration. Calculating the mean square displacement from
the reference configuration and making an integration over the strength of the
perturbation, it is possible to estimate the vibrational entropy \cite{noi}.
The difference $S-S_{vib}$ provides an estimate of the configurational entropy
$S_{conf}$ as a function of packing fraction $\varphi$ (or density $\rho$).

The main findings of the present work are the following.  
{\it i}) $S_{conf}$ is a decreasing function of packing fraction $\varphi$, 
and a suitable extrapolation to zero provides and estimate of the ideal 
phase transition point (Kauzmann packing fraction) 
$\varphi_{_K} \simeq 0.62$.
{\it ii}) The diffusivity
$D$ and configurational entropy $S_{conf}$ are related through the Adam-Gibbs
relation, in agreement with previous claims \cite{speedy3}.

\section{The model}

The studied system is a binary mixture $50-50$ of hard spheres, $A$ and $B$, with diameters
ratio $\sigma_{_B}/\sigma_{_A}$=$1.2$.  The collision diameters are
$\sigma_{_{AA}}$=$\sigma_{_A}$, $\sigma_{_{BB}}$=$\sigma_{_B}$ and
$\sigma_{_{AB}}$=$(\sigma_{_A}+\sigma_{_B})/2$.  The particles ($N$=$256$) are
enclosed in a cubic box with periodic boundary conditions.   We use the
following units: $\sigma_{_B}$ for length and $m_A=m_B=1$ for mass. Moreover
we chose $k_B=1$ and $\hbar=1$.  The density is measured by the packing
fraction $\varphi$ that is related to the number density $\rho=N/V$ by
$\varphi$=$\rho \pi (\sigma_{_A}^3+\sigma_{_B}^3)/12$. We analyzed systems in
the range $\varphi$=$0.425-0.580$. Hard spheres systems depend only trivially
on temperature that sets an overall scale for the dynamics, consequently we
perform all our simulations at $T=1$. We performed standard event-driven
molecular dynamics \cite{rapaport} and we stored several equilibrated
configurations at different density.
\begin{figure}[t]
\vspace{0.8cm}
\begin{center}
\includegraphics[width=.6\textwidth]{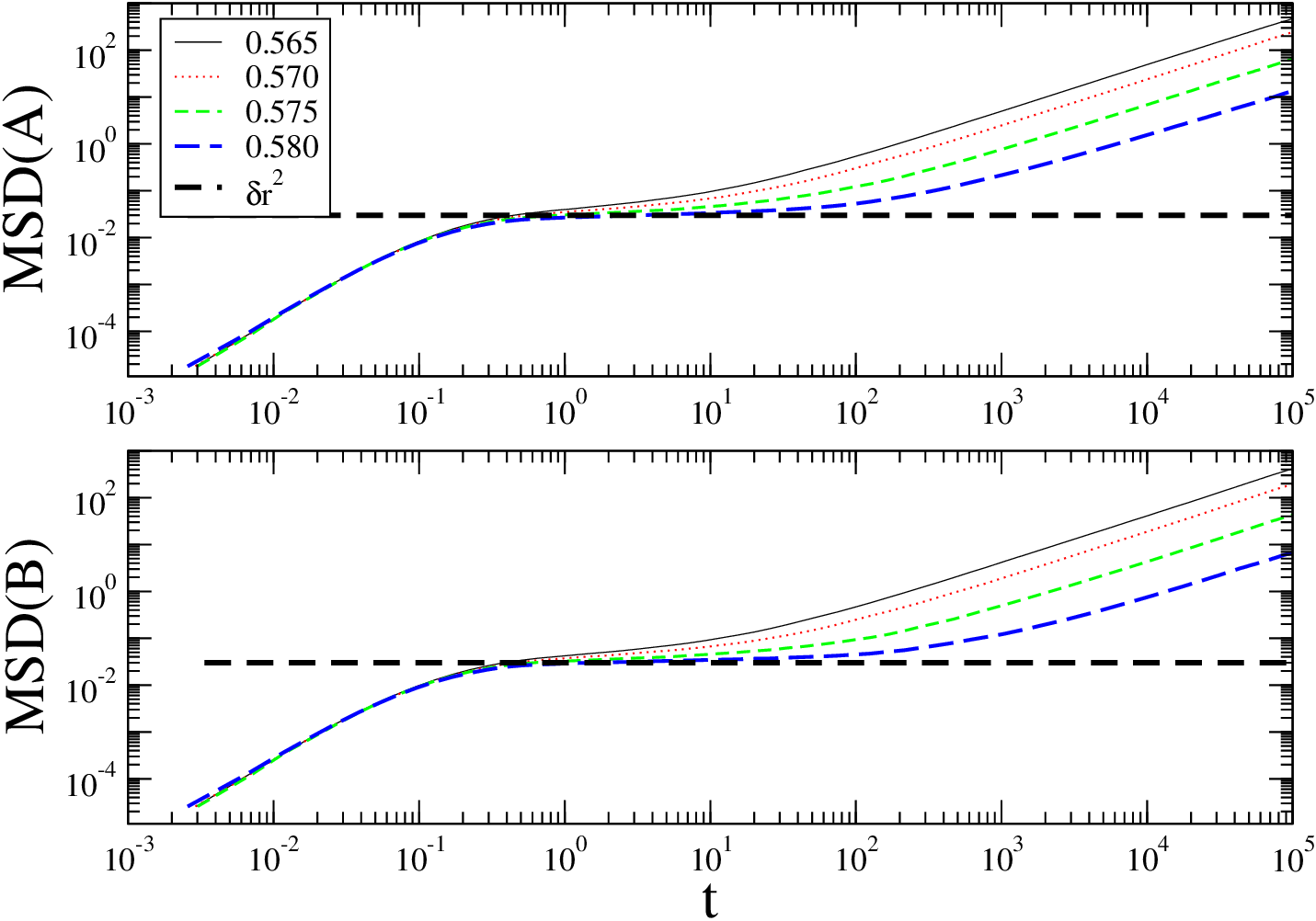}
\end{center}
\caption{ Mean square displacements (MSD) of the species $A$ (top) and
$B$ (bottom). Both MSDs has been normalized by the value of the
diameter squared.  The dashed lines represent the cage size squared,
$\delta r^2$.}
\label{fig1}
\end{figure}

\section{Diffusivity}

The diffusion coefficients $D$ of the two species have been extracted from the
long time limit of the mean square displacements (MSD) $\langle r^2(t)\rangle$=$N^{-1}\langle [{\bf
r}(t)-{\bf r}(0)]^2 \rangle$ (${\bf r}$ is the $3N$-vector of
the coordinates):
\begin{equation} 
\lim_{t\rightarrow \infty}\frac{\langle r^2(t)\rangle}{t}\simeq 6D \ .
\label{einstein} 
\end{equation} 
To improve the statistical significance of the data, an average over $10$
independent runs have been performed. 
In Fig \ref{fig1} the mean squared displacements for the slowest cases, 
i.e. $\phi >0.56$, are presented for both species.  
It is clear that on increasing the density the MSDs develop the typical 
two step relaxation pattern. The first part of the MSD is purely ballistic while, 
at later stage, it reaches the diffusive regime, described by Eq.~\ref{einstein}.  
Between this two regimes a plateau starts to develop. This is the clear indication 
of a caging effect. Each particle starts to feel the crowding of its neighbors and 
it is trapped in a cage for longer and longer time on increasing the density.
The height of the plateau is the typical ``cage diameter squared'', $\delta r^2$. 
For both species we find $\delta r^2 =3\cdot 10^{-2} \sigma_\alpha^2$ 
for $\alpha=A \mbox{ or } B$ represented by a dashed line in Fig.~\ref{fig1}. 
This is a clear evidence that the two species have the same caging effect. 
We shall return on the value of $\delta r^2$ later on.  

\begin{figure}[t]
\vspace{0.8cm}
\begin{center}
\includegraphics[width=.6\textwidth]{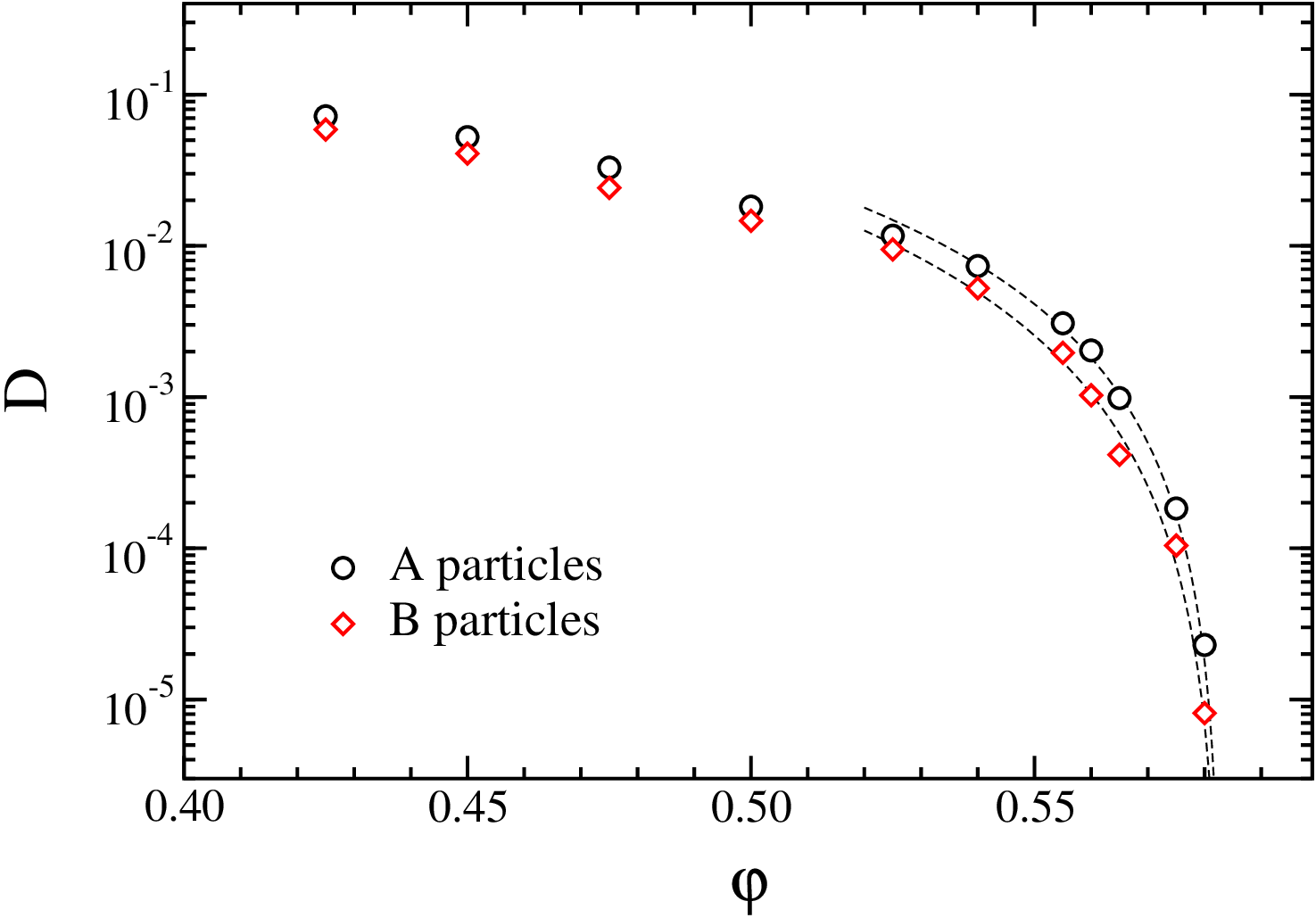}
\end{center}
\caption{
Diffusivity for A and B particles as a function of packing fraction $\varphi$.
The lines are  power-law fits for $\varphi \geq 0.53$,
$C\;(\varphi_c - \varphi)^\gamma$, 
with:
$\varphi_c$=$0.583$, $\gamma$=$2.27$, $C$=$9.50$ for A particles;
$\varphi_c$=$0.583$, $\gamma$=$2.47$, $C$=$11.66$ for B particles.
}
\label{fig2}
\end{figure}

In Fig. \ref{fig2} the diffusivities $D$ of A and B particles are
plotted as a function of the packing fraction $\varphi$. 
Dashed lines in the figure
are power-law fits $D = C\;(\varphi_c - \varphi)^\gamma$ of the high packing-fraction data 
($\varphi \geq 0.53$), 
as predicted by mode-coupling theory.
The fitted parameters are: 
$\varphi_c$=$0.583$, $\gamma$=$2.27$, $C$=$9.50$ for A
particles and $\varphi_c$=$0.583$, $\gamma$=$2.47$, $C$=$11.66$ for B
particles.  We note that both diffusivities give rise to the same
mode-coupling packing fraction $\varphi_c$, in agreement with the
prediction of the theory \cite{Voigtmann} and with previous
simulations of the same model \cite{foffietal}.

\section{Configurational entropy}

\begin{figure}[t]
\vspace{0.8cm}
\begin{center}
\includegraphics[width=.6\textwidth]{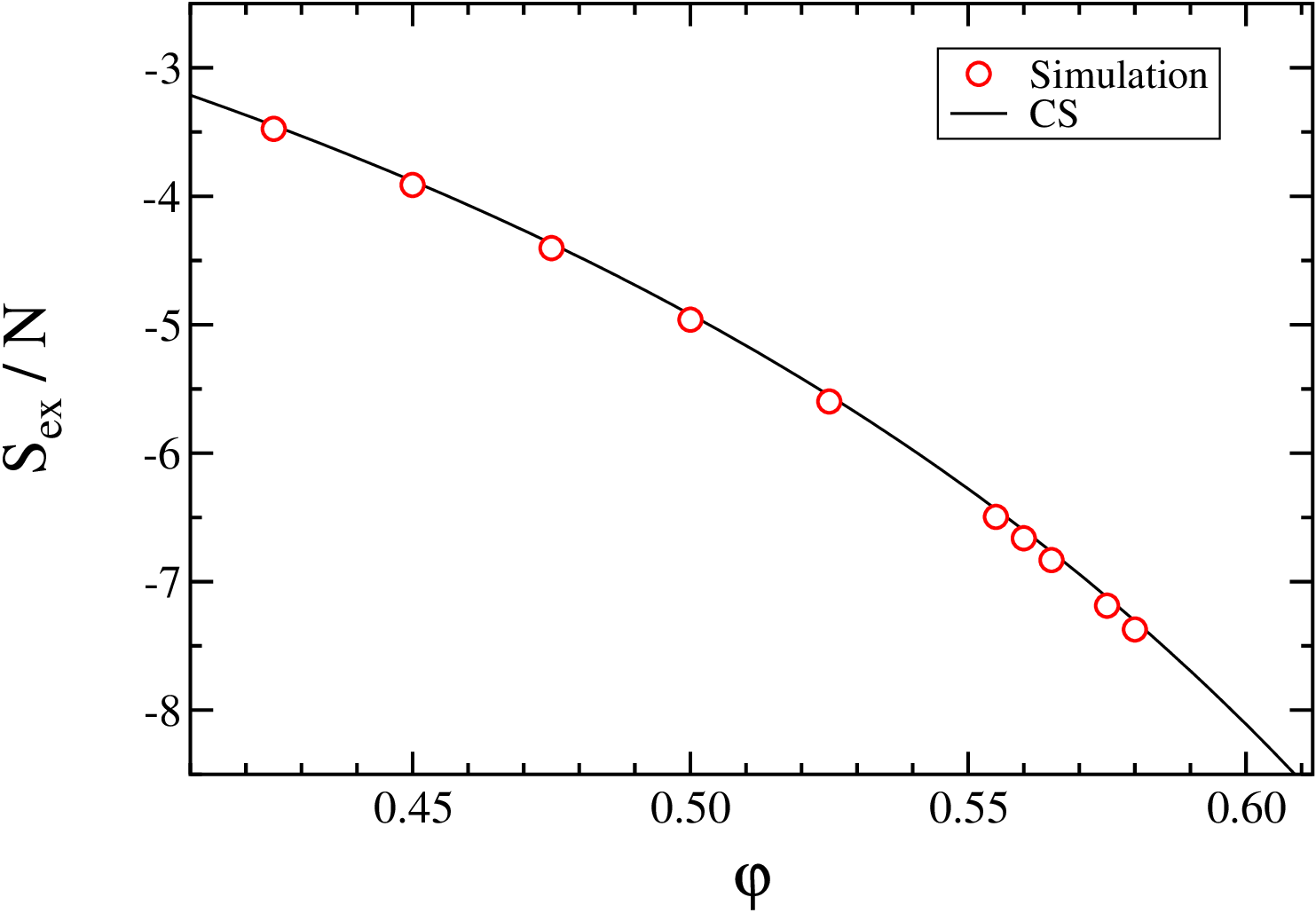}
\end{center}
\caption{
Excess entropy $S_{ex}$ for the mixture of hard spheres as obtained from our simulations
(symbols) compared with the analytical Carnahan-Starling expression (line) \cite{cs,nota_cs}.
}
\label{fig3}
\end{figure}

We now turn to the calculation of configurational entropy.  The method we
follow to estimate $S_{conf}$ requires the computation of the total entropy
$S$ and vibrational entropy $S_{vib}$.  The total entropy $S$ is calculated
{\it via} a thermodynamic integration from ideal gas and can be expressed as
\begin{equation}
S (\rho) = S_{id}(\rho) + S_{ex}(\rho) \ ,
\label{S2}
\end{equation}
where $S_{id}$ is the entropy of the ideal gas and $S_{ex}$ is the 
excess entropy with respect to the ideal gas.
For a binary mixture, the ideal gas entropy is:
\begin{equation}
\frac{S_{id}(\rho)}{N} = \frac{5}{2}-\ln{\rho}-3\ln{\lambda}+\ln{2} \ ,
\label{Sid}
\end{equation}
where $\lambda=(2\pi\beta\hbar^2/m)^{\frac{1}{2}}$ is the De Broglie
wavelength ($\hbar$ is the Planck's constant and has been set to unitary value),
and the term $\ln{2}$ takes into account the mixing contribution.
The term $S_{ex}$ can be expressed in the following form
\begin{equation}
S_{ex}(\rho)  = - \frac{N}{T}
\int_{0}^{\rho} \frac{d\rho}{\rho^2}\  P_{ex} \ ,
\label{Sex}
\end{equation}
with $P_{ex}$ the excess pressure. We extracted $P_{ex}$ from the zero density
limit up to the densities of interest, performing numerical simulations and
fitting the results of the pressure with a high order polynomial in $\rho$.
In Fig. \ref{fig3} we show the numerically calculated excess entropy $S_{ex}$
(symbols) together with the analytic estimate provided by the
Carnahan-Starling (CS) equation of state, extended to hard sphere mixtures
\cite{al,cs,nota_cs}.  We note that at high densities the CS equation of state
overestimates the entropy of about $7\%$. This discrepancy, however, is
not sufficiently significant to affect the resulting $S_{conf}$ values, in
particularly closed to the glass transition.

\begin{figure}[t]
\vspace{0.8cm}
\begin{center}
\includegraphics[width=.6\textwidth]{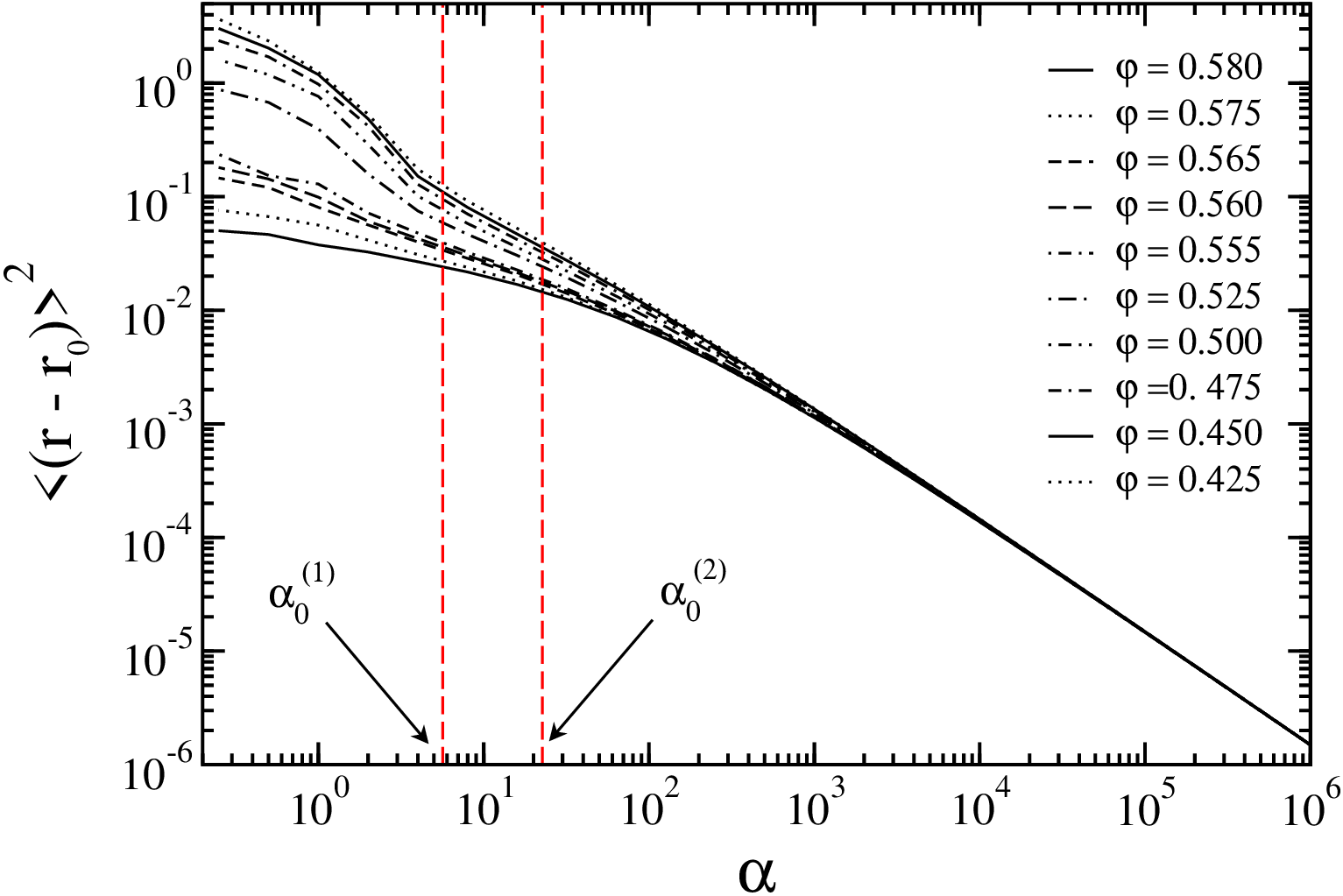}
\end{center}
\caption{
The quantity $\langle ( {\bf r} - {\bf r}_0 )^2 \rangle_{\alpha}$
plotted vs. $\alpha$ in logarithmic scale for different packing fractions
$\varphi$.
Vertical lines are the values  $\alpha_0^{(1)}=2^{2.5}$ and 
$\alpha_0^{(2)}=2^{4.5}$ used as $\alpha_0$-value in the integral 
in Eq.~\ref{svib2} for the calculation of $S_{vib}$.
}
\label{fig4}
\end{figure}

The method we use for the calculation of $S_{vib}$ is based on the 
investigation of a perturbed system
\begin{equation}
\beta H' = \beta H + \alpha N ({\bf r} - {\bf r}_0)^2 \ ,
\end{equation}
where $H$ is the unperturbed hard spheres Hamiltonian, $\alpha$ is the
strength of the perturbation, ${\bf r}_0$ specifies the particles coordinates
of a reference configuration and $({\bf r} - {\bf r}_0)^2 \equiv
N^{-1}\sum_{i=1}^{N} ({\vec r}_i - {\vec r}_{0,i})^2$.  The reference
configuration ${\bf r}_0$ is chosen from equilibrium configurations at the
considered density (randomly extracted from the stored configurations obtained
during molecular dynamics simulations).  With this choice one is sure that the
estimated vibrational entropy (see formula below) pertains to the correct
state at the studied density.  The vibrational entropy can be obtained from
the formula (see Ref. \cite{noi} for details):
\begin{equation}
\frac{S_{vib}}{N} = \int_{\alpha_0}^{\alpha_{\infty}} d\alpha' \langle ({\bf r} - {\bf r}_0)^2 \rangle_{\alpha'}
- \frac{3}{2} \ln\left(\frac{\alpha_{\infty} \lambda^2}{\pi}\right) + \frac{3}{2} \ ,
\label{svib2}
\end{equation}
where $\alpha_{0,\infty}$ are the lower/upper limit of integration, 
and $\langle ... \rangle_{\alpha'}$ is the canonical average 
for a given $\alpha'$.
The choice of $\alpha_0$ deserves few comments.  If the system were confined
to move inside a given local free-energy minimum, 
for a correct estimation of $S_{vib}$ one would take the
lower limit $\alpha_0$=$0$ in the integral in Eq. \ref{svib2}.  As the system,
at enough low value of $\alpha$, begins to sample different states (the
harmonic force due to the perturbation is no more able to constrain the system
inside one state), $\alpha_0$ has to be chosen in such a way that the system
has not yet left the state: the underlying idea is that Eq. \ref{svib2} gives
a correct estimation of $S_{vib}$ until the system remains trapped in the
state.  An appropriate choice in our case seems to be
$\alpha_0^{(1)}$=$2^{2.5}$  for all the densities 
, as, close to this point, one observes a crossover
for all the investigated densities (more pronounced for low density data).  In
Fig. \ref{fig4} the quantity $\langle ({\bf r} - {\bf r}_0)^2
\rangle_{\alpha}$ is reported as a function of $\alpha$. An arrow indicates
the chosen value $\alpha_0$=$\alpha_0^{(1)}$, below which one observes the
crossover associated to the exploration of different states.

It is worth noting that different choices of $\alpha_0$ are in principle
possible, giving rise to different estimations of the vibrational entropy
term.  However, even though a kind of arbitrariness is present in the method,
one can argue that a reasonable choice should be for values above
$\alpha_0^{(1)}$ (close to the crossover corresponding to the exploration of
different states) and below an upper value $\alpha_0^{(2)}$ at which one is
sure that the system is still confined in a single state.  The latter value
can be estimated requiring that the MSD $\langle ({\bf r} - {\bf r}_0)^2
\rangle$ is always close/below the cage diameter squared $\delta r ^2 \simeq
3 \cdot 10^{-2} \sigma_\alpha^2$ (with $\alpha=A \mbox{ or }B$) (this has been
estimated from the {\it plateau} of the mean square displacement, see
Fig.~\ref{fig1}).

The chosen value in our case is $\alpha_0^{(2)}$=$2^{4.5}$
(indicated by an arrow in Fig. \ref{fig4}).
We then repeated the same calculation of $S_{vib}$ using Eq.\ref{svib2} with 
the lower bound in the integral $\alpha_0=\alpha_0^{(2)}$.
In this way we obtain a lower and upper bound for the quantity of interest $S_{vib}$,
by using respectively $\alpha_0^{(1)}$ or $\alpha_0^{(2)}$ in the 
expression of $S_{vib}$ in Eq.\ref{svib2}.

\begin{figure}[t]
\vspace{0.8cm}
\begin{center}
\includegraphics[width=.6\textwidth]{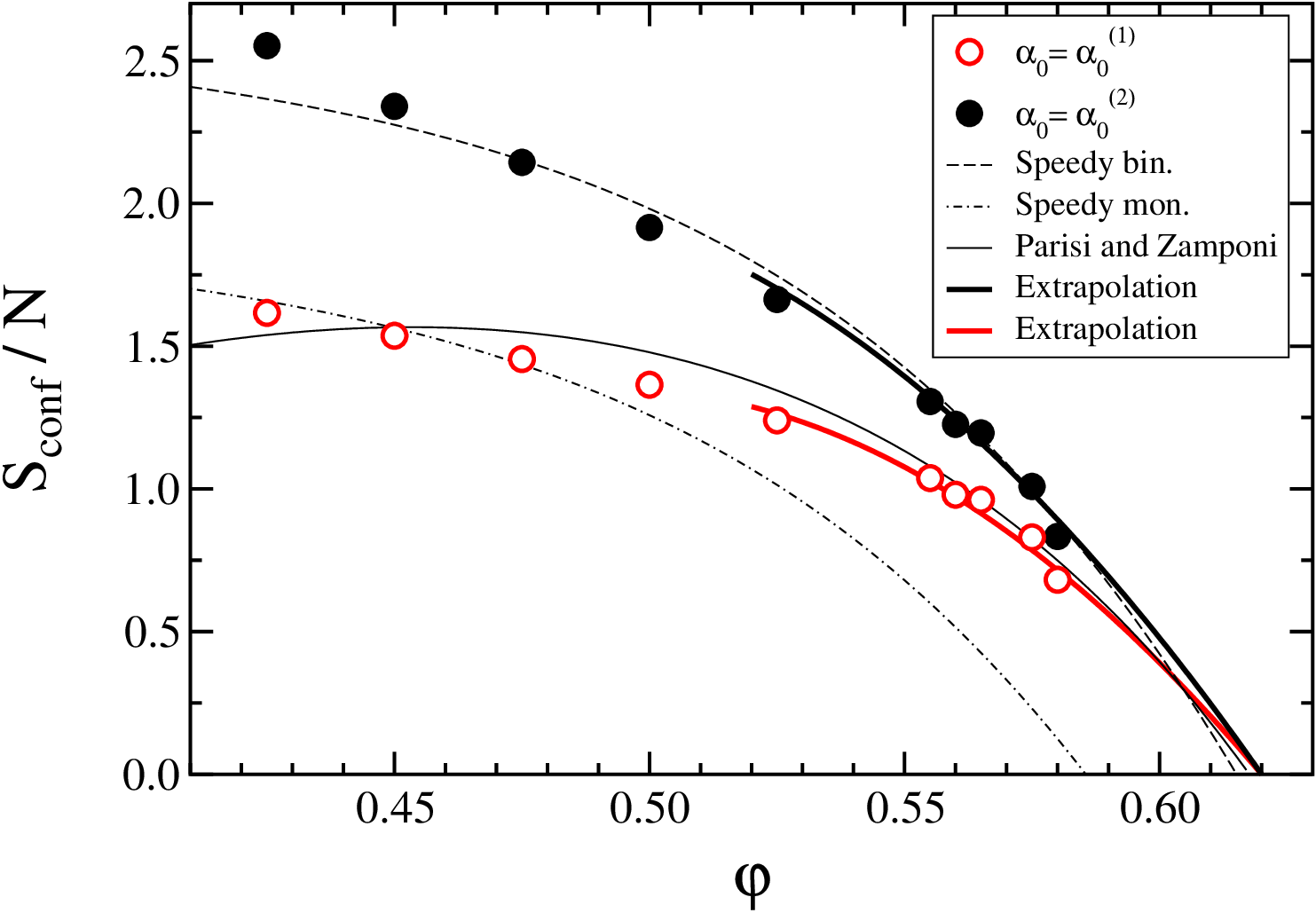}
\end{center}
\caption{ Configurational entropy $S_{conf}$ as function of packing fraction
  $\varphi$.  Open symbols are data using $\alpha_0^{(1)}=2^{2.5}$, full
  symbols using $\alpha_0^{(2)}=2^{4.5}$ (see text).  Dashed and dot-dashed
  lines are from Speedy \cite{speedy3} for binary and monatomic hard-spheres
  respectively.  Thin full line is the analytical computation of Parisi and
  Zamponi for monatomic hard-spheres \cite{paza}.  Thick lines are polynomial
  extrapolations of our data in the high packing fraction region, giving rise
  to the same Kauzmann packing fraction estimation for which
  $S_{conf}(\varphi_K)$=$0$: $\varphi_{_K} \simeq 0.62$.  }
\label{fig5}
\end{figure}

Fig. \ref{fig5} shows $S_{conf}$ as a function of $\varphi$. The
configurational entropy $S_{conf}$ is calculated using Eq. \ref{entropy},
where the two entropies $S$ and $S_{vib}$ are obtained from Eq.s \ref{S2} and
\ref{svib2} respectively.  Due to the fact that the correct integral for the
estimation of $S_{vib}$ should be done from $\alpha_0$=$0$, but with the
system always inside a given state, we have added to the expression in Eq.
\ref{svib2} the term $\alpha_0 \; \langle ({\bf r} - {\bf r}_0)^2
\rangle_{\alpha_0}$, corresponding to assume a constant value of $\langle
({\bf r} - {\bf r}_0)^2 \rangle$ below $\alpha_0$ and using a zero value for
the lower limit of the integral in Eq. \ref{svib2}.  Fig. \ref{fig5} shows the
two estimates of $S_{conf}$, corresponding to the two different values of
$\alpha_0$: $\alpha_0^{(1)}$=$2^{2.5}$ (open symbols) and
$\alpha_0^{(2)}$=$2^{4.5}$ (full symbols).  One observes that the discrepancy
between the two estimations decreases by increasing the packing fraction,
suggesting that, at high density, the method used to calculate
$S_{conf}$ is less affected by the choice of the free parameters entering in
its evaluation.  This is probably due to the fact that increasing the density
the system tends to be more trapped in a local free-energy minimum.
Indeed, it is only at high density that the method is expected to work better,
due to the better definition of two time scales corresponding to local-fast 
and global-slow dynamics (see Fig. 1). At low density, instead, the two are
less separated and this corresponds to a difficulty in the extrapolation for 
$\alpha_0 \to 0$ of the quantity reported in Fig. 4. The low density data 
show a  more clear crossover on lowering $\alpha$, and then
a worst definition of state in this limit. As we are interested in the high 
packing fraction extrapolation, this fact do not affect our prediction on the 
Kauzmann density value.
Also reported in the
figure are the curves obtained by Speedy \cite{speedy3} using a 
different method 
 (assuming a particular form of the vibrational entropy, a Gaussian 
distribution of states and involving some free parameters)
for the estimation of $S_{conf}$, for monatomic (dot-dashed
line) and binary (dashed line) hard-spheres (with the same diameters ratio
$1.2$ and composition $50:50$ as in our case).  
It is worth noting that our method improves on Speedy's one, as, 
even though requiring some accuracy in 
the choice of the $\alpha_0$ parameter, has the advantage to be less affected by 
the presence of many free parameters and particular assumptions.
We note that the data of
Speedy for the binary case do agree very well with our data with
$\alpha_0$=$\alpha_0^{(2)}$, suggesting the possibility that the choice of
$\alpha_0$=$\alpha_0^{(2)}$ is more accurate for the estimation of $S_{vib}$
and so of $S_{conf}$.  As a comparison, in Fig.~\ref{fig5} is also reported an
analytic estimation of $S_{conf}$ recently provided by Parisi and Zamponi
\cite{paza} for monatomic hard-spheres.  From the $\varphi$-dependence of the
configurational entropy one can determine the packing fraction at which
$S_{conf}$ extrapolates to zero, corresponding to the ideal phase transition
point (Kauzmann packing fraction $\varphi_{_K}$) $S_{conf}(\varphi_{_K})$=$0$.
Using a polynomial extrapolation \cite{nota_poly} 
for the two set of data (corresponding to the
different $\alpha_0$ values) we obtain an estimated Kauzmann packing fraction
value $\varphi_{_K} \simeq 0.62$ (see Fig.\ref{fig5}).
It is worth noting that, even
though the two curves 
are quite different, the estimated value of $\varphi_{_K}$ is the same, again
suggesting the robustness of the method in the high density region and then in
the estimation of the Kauzmann packing fraction.

\section{Adam-Gibbs relation}
\begin{figure}[t]
\vspace{0.8cm}
\begin{center}
\includegraphics[width=.6\textwidth]{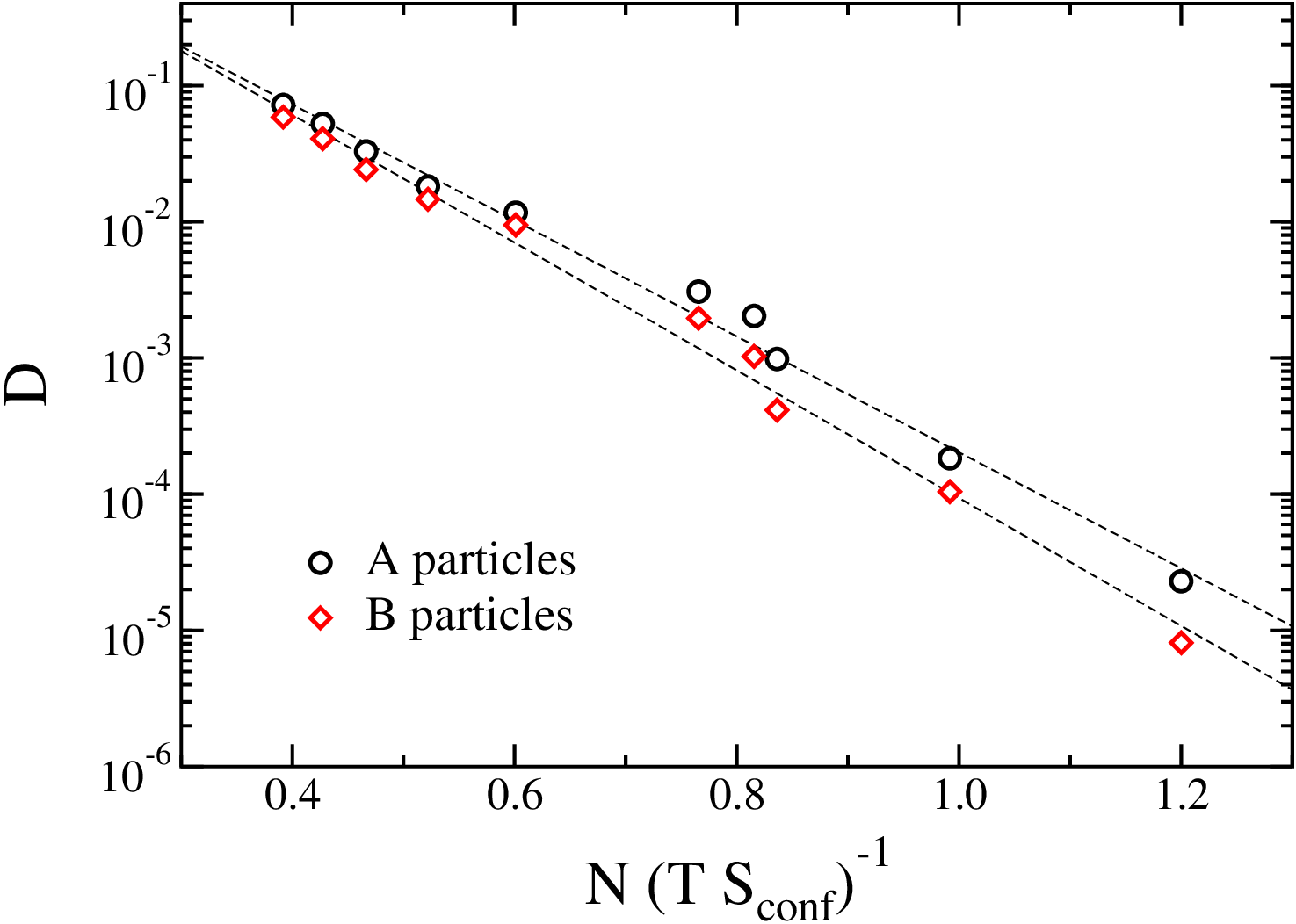}
\end{center}
\caption{The
Adam-Gibbs relation 
$D$=$D_{\infty}\exp[-N\Delta/(TS_{conf})]$ 
($T$=$1$)
for the two species A and B:
$D_{\infty}$=$3.66$, $\Delta$=$9.8$ for A particles;
$D_{\infty}$=$4.59$, $\Delta$=$10.5$ for B particles.
The data of $S_{conf}$ are obtained with $\alpha_0$=$\alpha_0^{(2)}$.
}
\label{fig6}
\end{figure}

In this Section we explore the validity of the Adam-Gibbs relation,
linking dynamic quantities, like diffusivity, 
to $S_{conf}$.
In Fig.~\ref{fig6} we report the diffusivities $D$
for A and B particles vs. 
the quantity $(TS_{conf})^{-1}$,
with $S_{conf}$ obtained for the value $\alpha_0$=$\alpha_0^{(2)}$.
We find that the Adam-Gibbs relation
\begin{equation}
D=D_{\infty} e^{-N\Delta/TS_{conf}}
\end{equation} 
is verified (lines in the figure), with: 
$D_{\infty}$=$3.66$, $\Delta$=$9.8$ for A particles;
$D_{\infty}$=$4.59$, $\Delta$=$10.8$ for B particles.
A similar behavior is obtained using $S_{conf}$ calculated with
$\alpha_0$=$\alpha_0^{(1)}$ (not shown in the figure), with the values:
$D_{\infty}$=$24.5$, $\Delta$=$9.5$ for A particles;
$D_{\infty}$=$37.5$, $\Delta$=$10.5$ for B particles, suggesting that, in this range of
diffusivity values,  the AG expression is not able to discriminate between the two different
estimations of $S_{conf}$.

\section{Conclusions}

In conclusion, we have calculated $S_{conf}$
for binary mixture hard spheres,
by numerically estimating the total entropy $S$ ({\it via} thermodynamic
integration from ideal gas) and the vibrational entropy $S_{vib}$  
using a numerical procedure based on Frenkel-Ladd method and recently applied
in the analysis of Lennard-Jones systems and attractive colloids:
the system is constrained inside a given ``{\it state}''
through an harmonic perturbed term in the Hamiltonian.
We found, in agreement with analytical and simulation results in the literature,
that $S_{conf}$ is a decreasing function of the packing fraction
$\varphi$, suggesting the possibility of a vanishing of $S_{conf}$ around
the Kauzmann point $\varphi_{_K}$=$0.62$.
Moreover, by studying the relationship between $S_{conf}$ and the diffusion
constant $D$, the Adam-Gibbs relation is found to reasonably hold for the analyzed system.

\ack
We thank G Ruocco, F Sciortino and F Zamponi for useful discussions and suggestions.
G F acknowledges the support of the Swiss Science Foundation (Grant No. 200021-105382/1).


\begin{thebibliography}{}
\bibitem{nota_cs}
The analytical CS expression for $S_{ex}$ is given by Eq. $9$ in Ref. \cite{cs}
omitting the last term $\ln Z$, as in our case the excess entropy is defined with respect
to the ideal gas at the same density instead of at the same pressure as in Ref. \cite{cs}
\bibitem{nota_poly}
We use a polynomial function of the form
$A\;(\varphi_{_K} - \varphi) - B(\varphi_{_K} - \varphi)^2$
to fit the data with $\varphi>0.5$. Adding more polynomial orders do not 
significantly affect the extrapolated Kauzmann packing fraction value.
\end{thebibliography}
\end{document}